# Collective Autoionization in Multiply-Excited Systems: A novel ionization process observed in Helium Nanodroplets

A. C. LaForge[1*], M. Drabbels[2], N. Brauer[2], M. Coreno[3], M. Devetta[4], M. Di Fraia[5], P. Finetti[6], C. Grazioli[6], R. Katzy[1], V. Lyamayev[1], T. Mazza[7], M. Mudrich[1], P. O'Keeffe[3], Y. Ovcharenko[8], P. Piseri[4], O. Plekan[6], K. C. Prince[6], R. Richter[6], S. Stranges[9], C. Callegari[6], T. Möller[8], and F. Stienkemeier[1]

[1]*Universität Freiburg, Germany;* [2]*EPFL Lausanne, Switzerland;* [3]*CNR-IMIP Rome, Italy;* [4]*University of Milan, Italy;* [5]*University of Trieste, Italy;* [6]*Elettra-Sincrotrone Trieste, Trieste, Italy;* [7]*European XFEL GmbH, Germany;* [8]*Technische Universität Berlin, Germany;* [9]*University of Rome "La Sapienza", Italy;* e-mail: [*]*aaron.laforge@physik.uni-freiburg.de.*

**Free electron lasers (FELs) offer the unprecedented capability to study reaction dynamics and image the structure of complex systems. When multiple photons are absorbed in complex systems, a plasma-like state is formed where many atoms are ionized on a femtosecond timescale. If multiphoton absorption is resonantly-enhanced, the system becomes electronically-excited prior to plasma formation, with subsequent decay paths which have been scarcely investigated to date. Here, we show using helium nanodroplets as an example that these systems can decay by a new type of process, named collective autoionization. In addition, we show that this process is surprisingly efficient, leading to ion abundances much greater than that of direct single-photon ionization. This novel collective ionization process is expected to be important in many other complex systems, e.g. macromolecules and nanoparticles, exposed to high intensity radiation fields.**

Atomic clusters are an ideal target for observing collective interactions since they bridge the gap between few-body and condensed phase systems. Observation of collective processes when clusters are irradiated by intense light sources has been reported for a wide variety of systems, for reviews see [1,2]. In atomic systems, a general metric to differentiate the regimes of light-matter interaction is the Keldysh parameter, $\gamma = \sqrt{\frac{IP}{2E_{pond}}}$, where $IP$ is the ionization potential and $E_{pond}$ is the ponderomotive energy, or average quiver energy of an electron in the light field. For nonperturbative interactions ($\gamma \leq 1$), tunneling ionization is dominant, while for perturbative interactions ($\gamma \gg 1$), multiphoton ionization is dominant. For atomic clusters in intense radiation fields, the formation of a nanoplasma where the entire cluster becomes ionized is found to be nearly independent of the Keldysh parameter. While for IR radiation ($\gamma \leq 1$), the nanoplasma is formed due to a strong resonant coupling of the cluster to the light's electric field, for VUV radiation ($\gamma \gg 1$), the plasma is formed by single and multi-photon ionization as well as heating by inverse bremsstrahlung [1]. Recently, the observation of nanoplasmas has even been extended into the x-ray regime [3].

Here, we present first experimental results on a novel collective ionization mechanism based on resonant excitation of clusters. Ionization occurs as any two neighboring excited atoms in the cluster transfer energy, leading to the decay of one atom and the ionization of the other. The process is shown schematically in Figure 1. This ionization mechanism is analogous to autoionization in atomic systems following double excitation of the system [4]. However, in contrast to the atomic case, the excitation occurs between many neighboring atoms, so the term Collective

Autoionization (CAI) provides a better description of the process. Such processes were recently proposed by Kuleff *et al*. [5]. Since the decay is dependent on the energy transfer between neighboring atoms, it is a type of Interatomic Coulombic Decay (ICD) initially predicted by Cederbaum et al. [6] and since then observed in several weakly bound systems, like neon clusters [7, 8], helium dimers [9] and water clusters [10,11].

Unlike conventional ICD [5], CAI is a resonant process where intense radiation is needed to excite at least two atomic partners. Kuleff *et al*. performed detailed calculations for neon clusters and found an enhancement of the ion production by two orders of magnitude compared to resonant two-photon ionization [5]. A similar enhancement was also theoretically predicted for a coupled system of hydrogen atoms [12].

Experiments on argon clusters [13] have revealed that non-resonant as well as resonant excitation below the cluster ionization potential results in plasma formation. However, at that time the ionization mechanism could not be identified and it was assumed that resonant two-photon ionization is the dominant process. As we will show in the following, CAI is much more efficient than resonant two-photon ionization and even more efficient than direct ionization by a single photon, at least in the case of He clusters. The calculations by Kuleff *et al.* [5], and initial results on Ne clusters [14], suggest that this is a general trend.

In first-order perturbation theory, the ionization rate is:

$$\Gamma = \sigma I^n$$

where $\sigma$ is the cross section, $I$ is the photon intensity, and $n$ is the number of absorbed photons. For photon energies below the ionization threshold, multiple photons are required to ionize the system, see e.g. [15, 16]. In contrast, ionization through CAI is based on the absorption of a single photon by multiple atoms within the cluster. The energy required for ionization of the excited atom is gained by the ICD of a neighboring excited atom. In the case of CAI one therefore expects a linear intensity dependence of the ionization rate. Since two-photon absorption by a single atom should have a quadratic power dependence, it should be possible to distinguish CAI from resonant two-photon ionization, at least in a certain intensity regime.

The experiment was performed at the low density matter end-station[17] at FERMI@Elettra [18,19] in Trieste, Italy and is described in detail in the Methods section. Helium nanodroplets, consisting of 50,000 atoms, were irradiated by VUV photons resulting in ionization of the constituent products. The photon energy was tuned either above or below ionization threshold. For excitation below ionization threshold, the photon energy was tuned to either to be resonant or nonresonant with the atomic energy level.

**Results**

The dependence of the ion signal on the light intensity is shown in Figure 2 for three photon energies, corresponding to three different ionization schemes. The normalized data is plotted on a log-log scale. As a result, the slope of a pseudolinear fit to the data is proportional to the number of absorbed photons, while the magnitude gives the relative ion abundance. At a photon energy of 42.8 eV, well above the ionization potential of helium, atoms are ionized by the absorption of a single photon. The slope of 1.07 ± 0.01 obtained from the fit clearly shows that direct ionization of the system is a one-photon process. At 20.0 eV, the clusters are transparent and the absorption of at least two photons is required for ionization. The slope of 2.06 ± 0.09 determined from the data confirms this. Since the 1s2p ← 1s$^2$ transition has the highest absorption cross section, it is the best candidate for observing CAI. For rare gas clusters, the absorption spectrum is quite different from atomic systems. The transitions are no longer narrow lines, but broad features shifted in energy with respect to the free atom due to the excited electron being perturbed by neighboring atoms [20]. As a result the maximum of the 1s2p ← 1s$^2$ absorption in the cluster corresponds to a photon energy of 21.4 eV [21]. For this resonant excitation of the system a slope of 0.63 ± 0.01 is found. The fact that the power dependence is less than one can be attributed to partial saturation of the 1s2p ← 1s$^2$ transition, which is estimated to have a large cross section, 25 Mbarn [22]. At the same time it suggests that the system is ionized via the collective ionization process described above.

**Discussion**

The most outstanding result is undoubtedly the very high ionization efficiency for resonant excitation considering that 2 photons are needed to ionize at this energy. For most of the FEL intensity range shown, the resonant helium-ion signal is almost an order of magnitude larger than that of direct ionization, a 1-photon process, and almost two orders of magnitude larger than that of the non-resonant case. A similar trend in the relative ion yields was also observed for dimer ions suggesting a general trend. Even qualitative considerations point out that resonant two-photon ionization is not sufficient to explain this observation, even assuming a full saturation of the excited level which is possible due to the large 1s2p ← 1s$^2$ resonant cross section, because the cross section for ionization from the excited 1s2p state, 0.02 Mbarn [23], is much smaller than that for direct 1-photon ionization(2.9 Mbarn at a photon energy of 42.8 eV [24]). Based on these considerations we conclude that collective autoionization of the excited atoms in the droplet is responsible for the large ion yield observed. This interpretation is supported by simulations.

Using a similar rate equation approach as Kuleff *et al*. [5] the ionization efficiency for CAI and resonant two-photon ionization has been calculated. The results of the simulations are presented in Figure 3 together with the experimental data. Overall, the calculated ion yields are in very good agreement with our experimental data with the introduction of a global scaling factor that takes into account experimental uncertainities and other factors (see Methods section). Over nearly the entire FEL intensity range, the ion yield is dominated by ions produced via CAI. Only at large intensities do ions produced via resonant two-photon ionization begin to contribute to the total ion yield. The simulations also show an excellent agreement with the data for direct ionization, thereby justifying the scaling of the laser intensity. The good agreement between the simulations and the experimental data corroborates the existence and efficiency of collective autoionization.

In conclusion, we observed a new ionization channel in clusters based on the collective autoionization of electronically excited atoms in the cluster. If the fraction of excited atoms is large, CAI becomes the dominant source of ion production with a rate much larger than that of direct ionization. Ionization by CAI is expected to be important for many other systems and of quite general character. With the advent of intense laser sources operating in the VUV regime, the possibility of exciting weakly-bound systems such as clusters, bio- and macro-molecules becomes more and more viable. When such systems are strongly ionized, a nanoplasma is formed which leads to its destruction. Furthermore, in systems involving species of atoms/molecules, CAI can also proceed through interspecies. Therefore, ionization by CAI is expected to be important to such system systems and of quite general character.

**Methods**

The experiment was performed at the Low Density Matter (LDM) end-station at the VUV seeded FEL FERMI@Elettra offering high tunability, narrow bandwidth (~ 20 meV), and high intensity (up to ~$10^{14} \frac{W}{cm^2}$) [18, 19]. The photon energies of the FEL were tuned via a seed laser and by setting the undulator gaps, and other machine parameters, according to predefined values. The FEL pulse length was assumed to be about 100 fs FWHM and the spot size was estimated to have a diameter 75 μm with a uniform spatial profile. The laser pulse energy at the setup is calculated from the value measured upstream on a shot-by-shot basis by gas ionization and the calculated nominal reflectivity of the optical elements in the beam transport system. The helium cluster beam was produced by supersonic expansion of helium at a backing pressure of 50 bar through a pulsed 50 μm nozzle that was cryogenically cooled to 18 K. From these expansion conditions and scaling laws, the helium cluster size was determined to be approximately 50,000 He atoms [25]. The cluster beam was perpendicularly crossed by the FEL beam at the center of the first acceleration region of a Wiley-McLaren time-of-flight mass spectrometer [26].

We use rate equations to simulate the power dependences in a similar fashion as Kuleff *et al*. [5]. In consideration of the observed strong ionization efficiency, it is assumed in the simulation that autoionization occurs between every excited pair of atoms that isn't ionized via resonant two-photon ionization. Furthermore, no secondary ionization processes were considered in the simulation. For the FEL beam, we assume a uniform spatial and a Gaussian temporal beam profile. Furthermore, we use the relevant excitation/ionization cross sections (direct ionization at hυ = 42.8 eV: 2.9 Mbarn; $1s2p \leftarrow 1s^2$ absorption cross section: 25Mbarn; ionization from the 2p state at hυ = 21.4 eV: 0.02 Mbarn). In the order to match the simulations to the experimental data, the estimated FEL pulse energy used in the simulation is reduced by a factor of 7.8. We attribute these losses to the additional beamline transmission losses and focus properties, nonuniformity in the spatial beam profile, and some uncertainties in our calculation of the $1s2p \leftarrow 1s^2$ absorption cross section and the use of an atomic cross section for ionization from the 1s2p state.

**Acknowledgements**

We would like to thank K. Ueda, L. Cederbaum, A. Kuleff, and P. Demekhin for their discussions on this subject and gratefully acknowledge the support of the staff of FERMI@ELETTRA.

**Author contributions**

A.C.L., M.Dr., M.De., R.K., V.L., Y.O., K.C.P., C.C., T.Mö., and F.S. designed and setup the experiment. A.C.L.,M.Dr., N.B., M.C.M.De. M.D.F., P.F. C.G. R.K. V.L., T.Ma., P.O.K., Y.O., P.P., O.P., K.C.P., R.R., S.S. C.C. T.Mö, and F.S. carried out the experiment. A.C.L. analyzed the data. A.C.L, M. Dr., and Y.O. performed the simulation. A.C.L., M.Dr., P.O.K., K.C.P., T.Mö., and F.S. interpreted the data. A.C.L., M.Dr., T.Ma., M.M., P.O.K., K.C.P.,C.C.,  T.Mö., and F.S. prepared the manuscript.

**Additional information**

The authors declare no competing financial interests. Reprints and permissions information is available online at http://npg.nature.com/reprintsandpermissions. Correspondence and requests for materials should be addressed to A.C.L.

Figure Captions

Figure 1. Process of collective excitation and collective autoionization (CAI). Intense FEL light interacts with multiple atoms within the cluster a) leading to collective excitation (excited atoms shown in red while ground state atoms are shown in blue) b). Through relaxation by CAI, some atoms transfer their excitation energy to neighboring atoms in an ICD process, thereby ionizing them while decaying to their ground state c) [5].

Figure 2. Power dependence and relative ion abundances for photon energies: 21.4 eV (black circles), 42.8 eV (red circles), 20.0 eV (blue circles) along with power dependence fits (lines of corresponding color).

Figure 3. Energy dependence of the relative ion abundances for photon energies: 21.4 eV (black circles) and 42.8 eV (red circles) scales on the bottom and left. Ion rate simulations corrected for transmission losses are plotted with the relative scales on the top and right. Solid line-total ion yield; dashed line-ions produced via CAI-dotted line: ions produced via resonant two-photon ionization (R2PI).

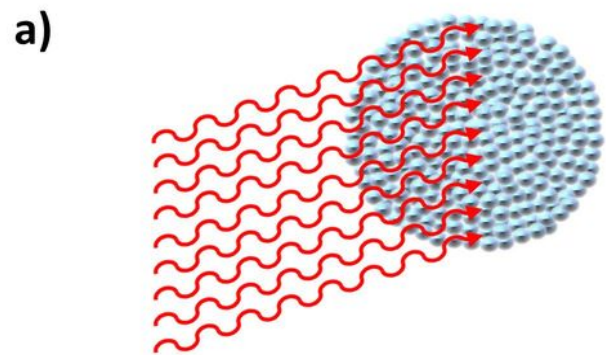
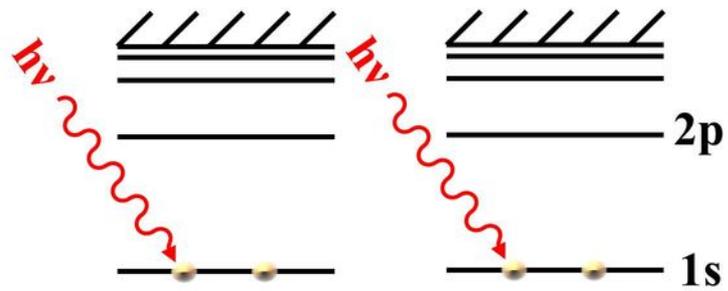
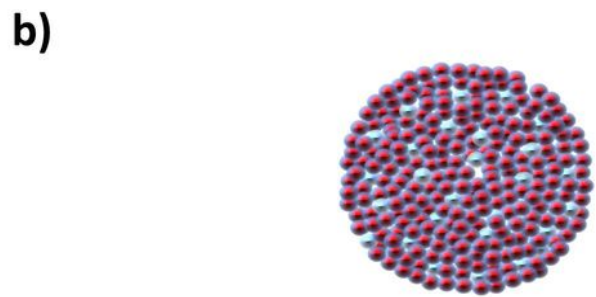
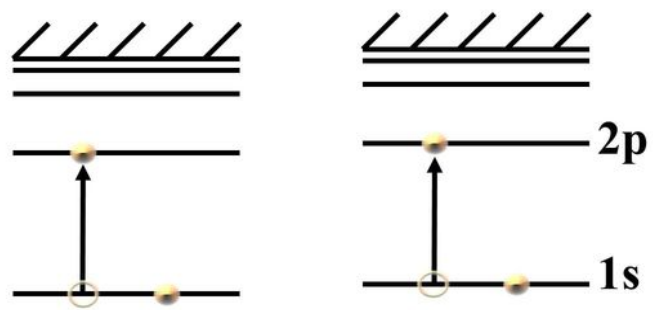
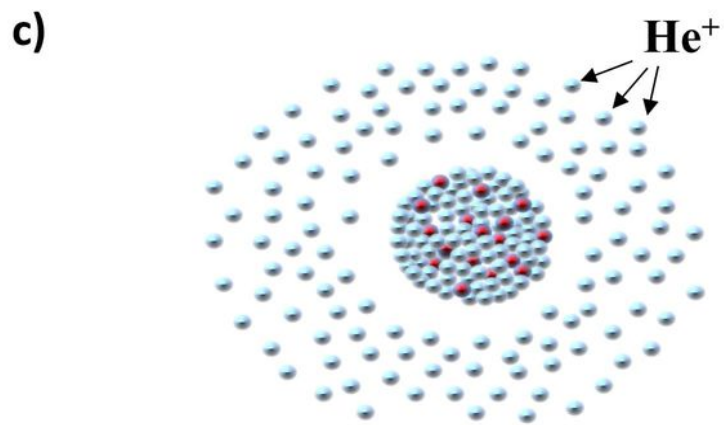
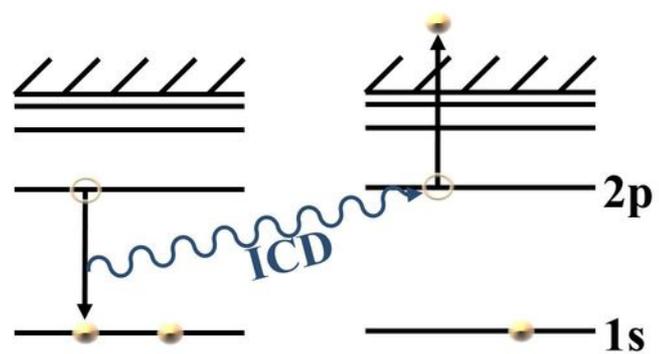

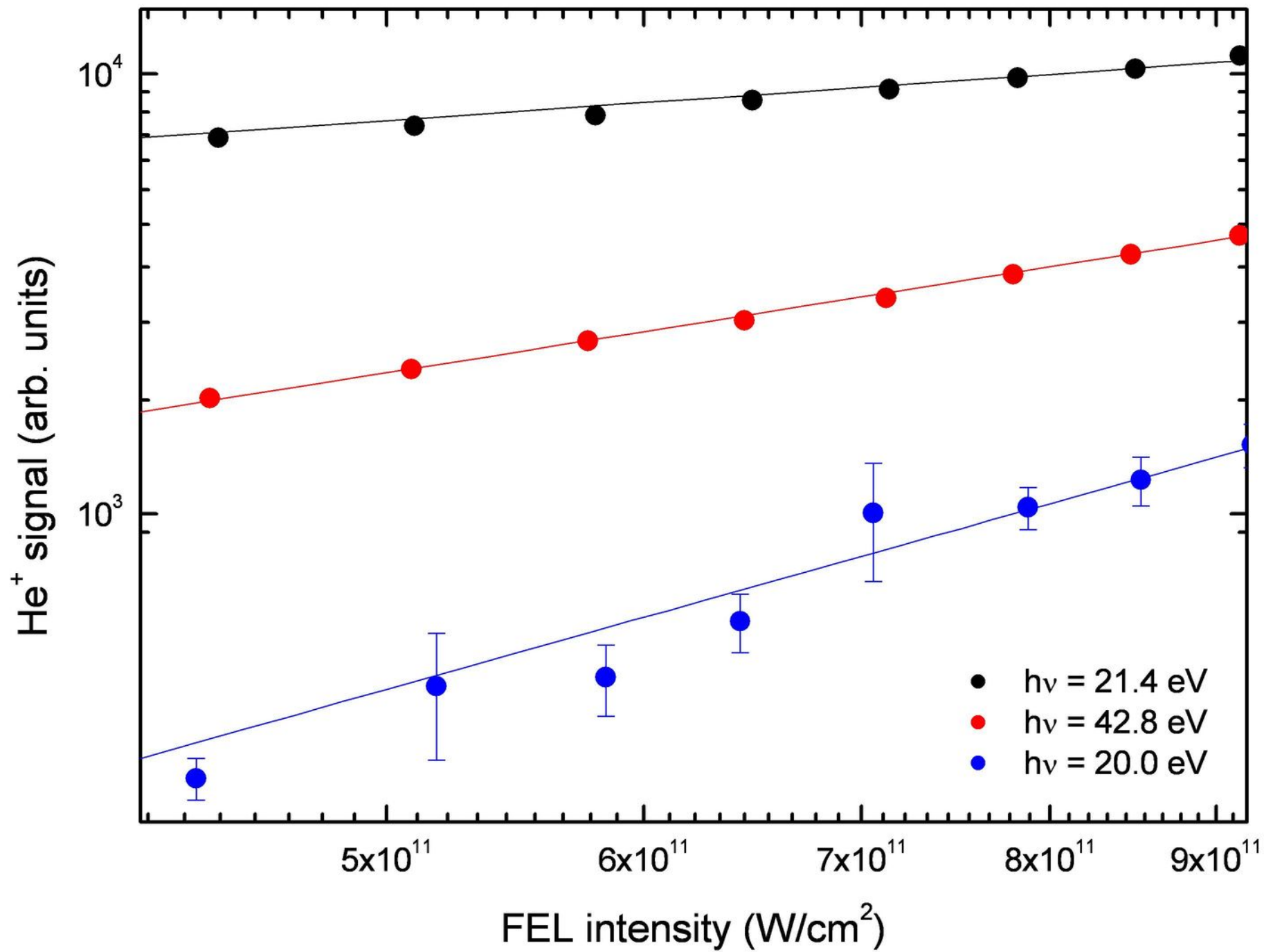

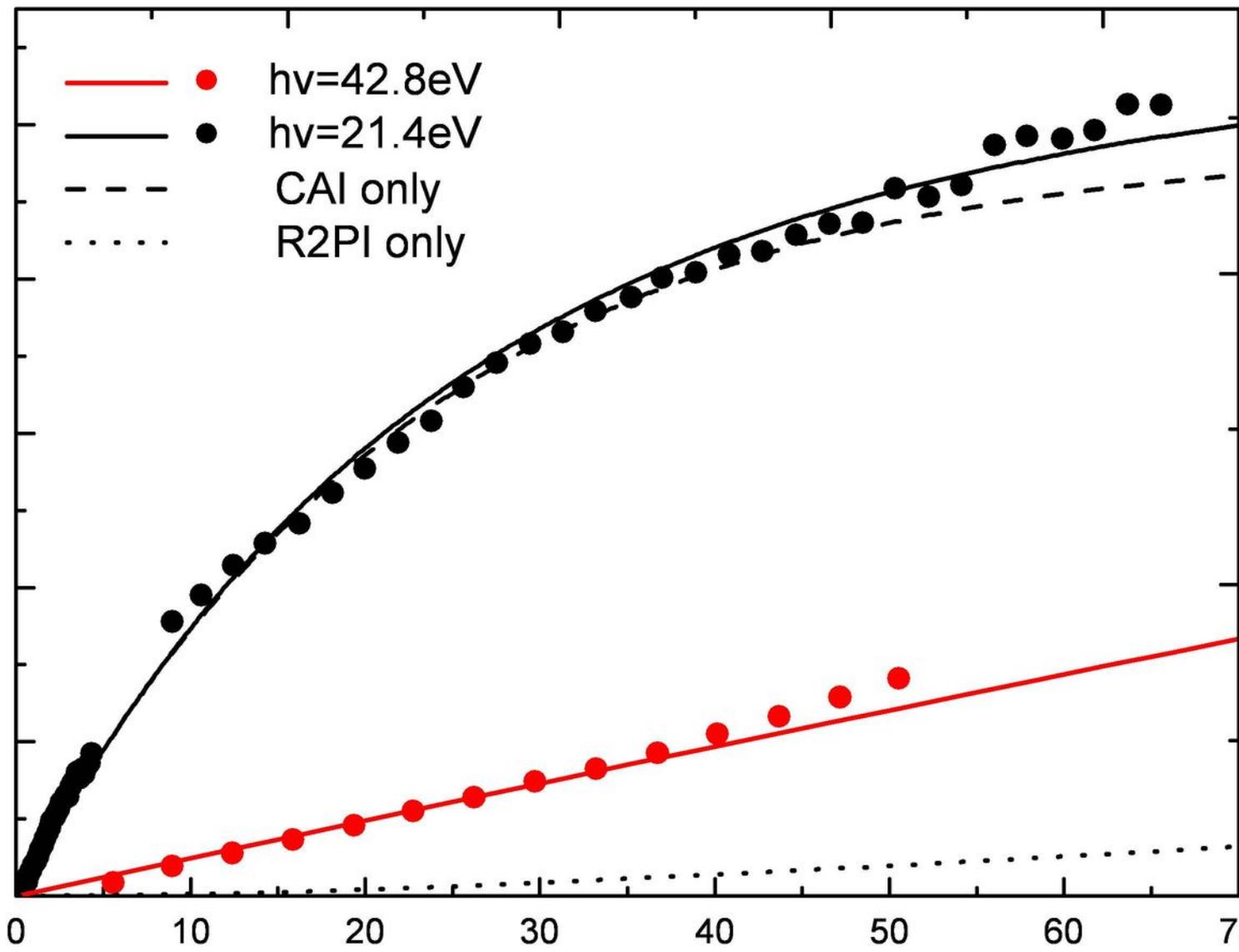